\documentclass[prb,amssymb,aps]{revtex4-1}
\usepackage{graphicx}
\usepackage{amsmath,amssymb}

\begin{document}
\draft
\title{Weinberg angle, current coupling constant, and mass of particles as
properties of culminating-point filters $-$ consequences for particle
astrophysics}
\author{E. Donth}
\address{Institut f\"{u}r Physik, Universit\"{a}t Halle, D-06120 Halle (Saale), Germany\\
E-mail: donth@physik.uni-halle.de}

\begin{abstract}
Culminating-point filter construction for particle points is distinguished
from torus construction for wave functions in the tangent objects of their
neighborhoods. Both constructions are not united by a general manifold
diffeomorphism, but are united by a map of a hidden conformal $S^{1}\times
S^{3}$ charge with harmonic (Maxwell) potentials into a physical space
formed by culminating points, tangent objects, and Feynman connections. \
The particles are obtained from three classes of eigensolutions of the
homogeneous potential equations on $S^{1}\times S^{3}$. The map of the $u(2)$
invariant vector fields into the Dirac phase factors of the connections
yields the electro-weak Lagrangian with explicit mass operators for the
massive leptons. The spectrum of massive particles is restricted by the
small, manageable number of eigensolution classes and an instability of the
model for higher mass values. \ This instability also defines the huge
numbers of filter elements needed for the culminating points. Weinberg
angle, current coupling constant, and lepton masses are calculated or
estimated from the renormalization of filter properties. Consequences for
particle astrophysics follow, on the one hand, from the restriction of
particle classes and, on the other hand, \ from the suggestion of new
particles from the three classes e.g. of dark matter, of a confinon for the
hadrons, and of a prebaryon. Definitely excluded are e.g. SUSY
constructions, Higgs particles, and a quark gluon plasma: three-piece
phenoma from the confinons are always present.\newline

{\bf Keywords:} 1. cosmological applications of theories

\ \ \ \ \ \ \ \ \ \ \ \ \ \ \ \ \ \ 2. dark matter

\ \ \ \ \ \ \ \ \ \ \ \ \ \ \ \ \ \ 3. classical tests of cosmologies

\ \ \ \ \ \ \ \ \ \ \ \ \ \ \ \ \ \ 4. baryon asymmetry\newline

\newpage
\end{abstract}

\maketitle

\tableofcontents
\newpage

\section*{1. Introduction}

A paradigm change in astroparticle physics is often not believed
to be necessary or even possible by most colleagues discussing the
progress in confirmation of the experimentally accessible part of
the Standard Model of particle physics. Moreover, two sociological
problems bother the communities. (1). The information between the
old and a new paradigm breaks down (Kuhn), if the new one contains
too many new conceptions and new links between them. (2). The
power of communities is increased with the number of active
members and with their intention to solve the problems during
their time. The second problem forms a sharper hierarchy
occasionally supporting
ignorance and even arrogance against the bare announcement of a new paradigm.%
\newline
\newline

To avoid Kuhn's breakdown of information, first paradigmatic attempts should
start from one or only a few key concepts which can understandably be
explained. Suppose that a new paradigm with many new concepts and links
truly exists which is consistent with the confirmed part of Standard Models.
Then, to neutralize the negative power of the communities, the key concepts
could be followed by a plausible, intuitive way to the new paradigm as a
whole.\newline
\newline

The physical key concept of the first attempt (this paper, later referred to
as \cite{this work}) is the difference between mass and energy behind the $%
E=mc^{2}$ formula. The key concept of a second attempt (referred to as (\cite%
{second part})) is a thermodynamic analysis of an initial liquid hypothesis:
Two phenomena known from the dynamic heterogeneity in the dynamical glass
transition are supposed to exist also in an initial liquid before the
cosmological inflation: $G$ defects (Glarum Levy defects) as seeds for later
galaxies, and $F$ speckles (Fischer speckles) claiming the largest causal
regions forming a finite universe.\newline
\newline

The mathematical tool of the first attempt is the filter \cite{Faure1964}
(section 2). Pedagogically, we may start in the Introduction from an
Euclidean space $E$ and ask about the physical use of its points ($x$). In
the frame of physical elementarity, these points $x$ are to locate
''pointlike'' particles in $E$, e.g. electrons. In quantum theory, the
particles have signatures (charge, spin, parity, ...) that are taken from
the wave function $\psi (x)$ in the neighborhood. Additionally, the
particles at $x$ have numerical parameters (mass, coupling constant,
Weinberg angle...) which are not taken from $\psi $: they are, from the
beginning, parameters in the wave equation determining $\psi $ as a function
of $x$. In such terms, the difference between mass and energy is physically
noted as $M/E$ separation and geometrically as $x/\psi $ separation.\newline
\newline

Compare the $x/\psi $ separation with a general mathematical topological
space of a set $E$. Then each point $x$ of $E$ is connected with a filter $%
{\frak F}(x)$; all the elements of ${\frak F}(x)$ contain the point $x$ and
are neighborhoods of $x\in E$. {\it Our physical space is therefore not such
a topological space with its identification of neighborhoods and filters},
because these two get different functions: The wave function $\psi $ belongs
to the neighborhood that cannot give the parameters of the particle. These
parameters, however, may belong to a filter ${\frak F}$, whose limit to some
culminating point can be the basis for their calculation. We could e.g. use
some kind of a Cauchy filter: having elements that can be assigned to
arbitrary small diameters needed for a culminating point construction of
physical point particles. We need, therefore, a physical space construction,
that distinguishes between the neighborhoods as $\psi $ carriers (later
called tangent objects), and the points as culminating points (whose filters
allow the calculation of parameters). This construction excludes the general
use of a space manifold with its metric diffeomorphism as e.g. used for the
gauge theory prior of the standard model.\newline
\newline

The preliminary stage of the above attempts \cite{this work,second part}, of
course, does not allow an exact prediction of the parameters for particle
astrophysics. In attempt \cite{second part}, only a qualitative, plausible
sketch of cosmological development before the primordial stage is tried. $-$
The suggested ''hidden charge model'' of \cite{this work} (shortly, the
model) indicates a new theoretical way to the parameters of the Standard
Model for elementary particles, especially to the reason why the model is so
as observed. Moreover, the way leads to some consequences beyond the
Standard Model physics. The stock of particles is strongly restricted, and
new phenomena become plausible (dark matter in \cite{this work}, and primary
flatness instead of dark energy, a hadronic inflation, a secondary role of
gravitation, and a phenomenological, thermodynamic reason for the desired
primordial fluctuations in \cite{second part}).\newline
\newline

The details and literature of the cosmological revolution can be learned
from the reports of the Particle Data Group \cite%
{Hagiwara,Eidelman,Yao2006,particle 4a} with their authentic, rational
reviews, the methods for the establishment of cosmological parameters from
Spergel's et al reports \cite{Spergel2003,Spergel2007}, and the development
of cosmology from \cite{Liddle,Perkins}. The main work on the hidden charge
model is published before the cosmological revolution: A relevant analysis
of the Spin Statistics Theorem \cite{Donth1970}, the hidden charge model %
\cite{Donth1986,DonthLange}, the construction of tangent objects \cite%
{Donth1988}, the analysis of the space problem \cite{DonthWZ89}, the
estimation of some parameters \cite{DonthWZ265}, and the classification of
eigensolutions of the hidden charge \cite{Busse1998}. A systematic, detailed
discussion of the whole construction is in a report \cite{Donth2006}, to be
published later.

\section*{2. Filters vs neighborhoods}

\subsection*{2.1 Filters}

The mathematics of filters (\cite{Faure1964} chapter 8 there) starts from a
{\it filter basis}. This is a non-empty family ${\frak B}$ of subsets of a
set E, if 1) the intersection of two subsets belonging to the family ${\frak %
B}$ also belongs to the family ${\frak B}$, and 2) the empty set
does not belong to the family ${\frak B}$. Examples are a set of
rectangles, all containing the point $A$ (figure~1a); or all sets
containing twodimensional disks $D^{2}$ with the center $A$
(figure~1b). If ${\frak B}$ is a filter basis on $E$, then the set
of all subsets of $E$ that contain at least one of the sets of
family ${\frak B}$, is called a {\it filter} (${\frak F}$)
generated by the basis ${\frak B}$ on $E$. In the example of
figure~1a, the basis generates a filter consisting of all sets of
rectangles containing the point $A$ (the later culminating point).

\begin{figure}[here]
\begin{center}
\includegraphics[width=0.5\linewidth]{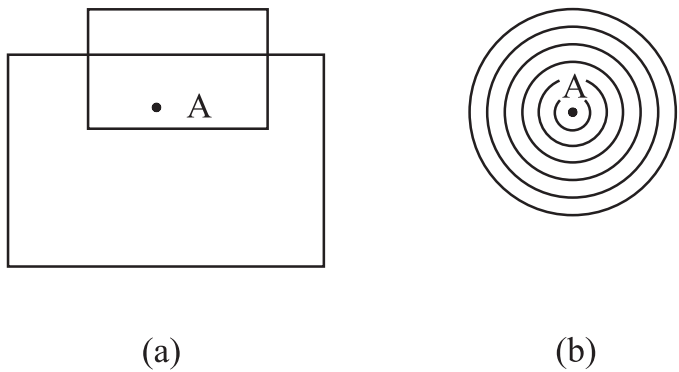}
\caption{Examples for filter constructions (details see text)
\label{fig1}}
\end{center}
\end{figure}

Three Remarks should help to sort our physical model into such mathematics:%
\newline

1. A {\it topological space} is a set $E$ each element $x$ (point $x$) of
which is connected with a filter ${\frak F}(x)$; all sets of these filters
must contain the point $x$, i.e. they are neighborhoods of the point $x$. In
addition, there is an analogeous condition for the neighborhoods \cite%
{Faure1964}. This filter ${\frak F}(x)$ is called filter of neighborhoods of
the point $x$.\newline
\newline

2. Be $f(x)$ a map of a set $E$ into a topological space $F$, and ${\frak F}$
a filter on $E$. A point $y\in F$ is called {\it limit of the map }$f${\it \
via the filter }${\frak F}(x)$ ($y=\lim\limits_{{\frak F}}f(x)$) if the
image ${\frak B}$ of the filter ${\frak F}$ gets its culminating point $y$
by the map $f$. A filter on a separable metric space is called {\it Cauchy
filter}, if it contains sets of arbitrary small diameters; a Cauchy filter
is a converging filter. A Cauchy filter has at most one culminating point.
If it has one, so the filter converges to this point.\newline
\newline

3. Filters that are related to the set of complements to natural numbers ($n$%
) are called Fr\'{e}chet filters. A sequence of points ($x_{n}$) is a Cauchy
sequence, if for any $\varepsilon >0$ exists such an integer $N$, that the
inequalities $n_{1}>N$, $n_{2}>N$ imply the inequality for the distance $%
\rho (x_{n_{1}},x_{n_{2}})<\varepsilon $. The Fr\'{e}chet filter associated
with $x_{n}$ is, therefore, a Cauchy filter.\newline
\newline

\subsection*{2.2 Charge model}

What are the hidden things that connect, in the sense of elementarity, the
culminating point $x$ with the wave function $\psi $? For this purpose, the
existence of a hypothetical {\it hidden charge} is assumed (figure 2): (1)
The hidden charge is an omnipresent compact ''electrodynamic'' $S^{1}\times
S^{3}$ manifold in four dimensions (geometrically called $H$ space; if
completed by an $A$ potential: hidden charge). (2) Homogeneous wave fields
for 1-form $A$ potentials on $H$ from Maxwell's equations (harmonic
potentials) are assumed. This construction is conformally invariant, i.e.
the coordinates are angles and do not have the meaning of lengths and their
''wave metric'' has no constant $c$ in m/s. The omnipresent hidden charge
suits therefore for the whole universe as well as for the particles therein,
from $R\approx 10^{-35}$m to $R\approx 10^{+26}$m, say. (Bell's theorem
excludes only local hidden parameters.) (3) Three torus coordinates ($\tau
,\varphi _{1},\varphi _{2},0...2\pi $) can carry potential components ($%
A_{0},A_{1},A_{2}$) whereas the one coordinate segment ($\vartheta ,0...\pi
/2$) does not carry a potential component, $A_{\vartheta }\equiv 0$. Any
kind of electrical charges on the $H$ space are excluded by div$_{4}A=0$.
(4) Culminating points are generated by an existential instability. The many
elements needed for the filters of culminating points come from the $%
\vartheta $ segment and are called vacuum elements: $z=\cos
^{2}\vartheta $ (or $\zeta =1-z=\sin ^{2}\vartheta $). (5) The $A$
potentials on the torus coordinates are mapped on the wave
functions $\psi $ in the tangent objects, the parameters of the
points $x$ come from the $\vartheta $ filter. (6) The culminating
points and their tangent objects form the physical or reality
space $M$, the filter belongs to the $H\longmapsto M$ map. The
whole construction is called (hidden) charge model.

\begin{figure}[here]
\begin{center}
\includegraphics[width=0.5\linewidth]{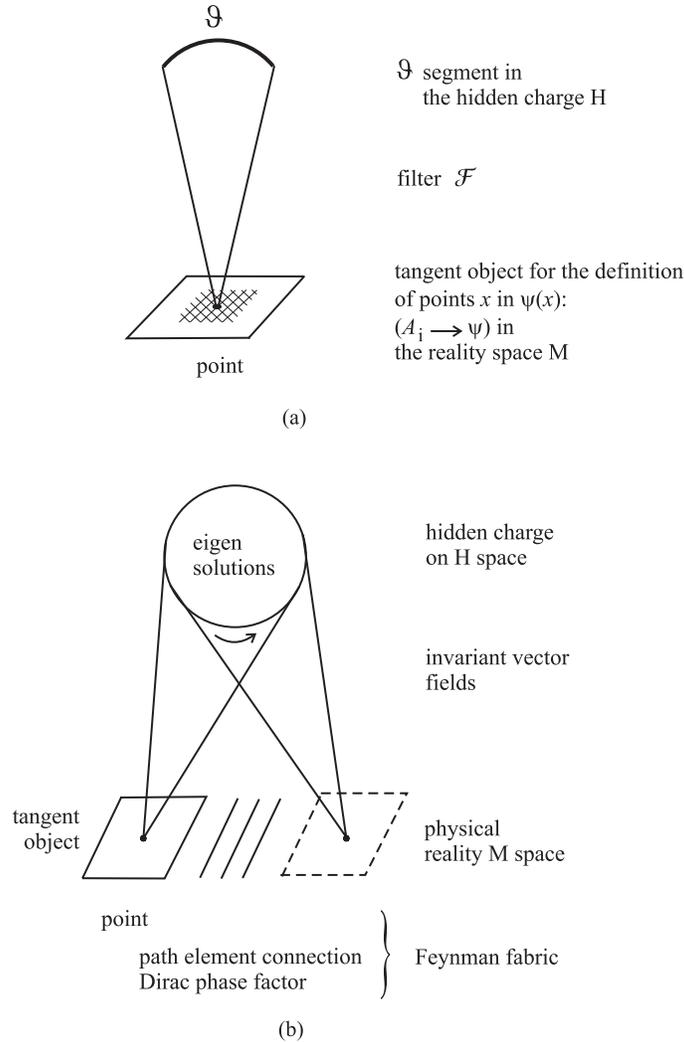}
\caption{Illustration of two methodical steps for the $x/\psi $
separation. (a). Culminating point and tangent object. (b).
Feynman fabric of points and connections over tangent objects. The
set of reality $M$ space elements \{points, tangent objects,
connections\} forms a ''crumbled space'' (section 4.1, below).
This space is not so ''tight'' as a manifold. \label{fig2}}
\end{center}
\end{figure}

The physical picture resulting from the model is not complicated. We can use
prosy and heuristic {\it physical} arguments: It is not necessary first to
be lost in the abyss of deep new mathematics.\newline
\newline

The culminating point / tangent object: the $x/\psi $ separation of the $%
H\longmapsto M$ map is illustrated by two methodical steps. First, the
filter of the $\vartheta $ segment vacuum elements is depicted by a segment
cone of the $H\longmapsto M$ map. This is e.g. for mass at $x$ (figure 2a).
Second, motions in the reality space $M$ are generated by ''rotations''
(invariant vector fields) of the hidden\ $H$ space giving $\psi $. This
leads to Dirac phase factors for Feynman path integrals of the electroweak
interaction. This is e.g. for energy (figure 2b, cf. also section 4).\newline
\newline

The key question for the $x/\psi $ separation is: What is the $x$
in the wave function $\psi (x)$, if there is no manifold for $x$?
We need (1) some construction, where $x$ can get a firm geometric
base that would be measurable by e.g. photons. This is the tangent
object around any point $x$. We need (2) some construction that
defines $\psi $. This is the ''Feynman fabric'' of random points
and Dirac phases between (figure 3) defining path integrals.

\begin{figure}[here]
\begin{center}
\includegraphics[width=0.5\linewidth]{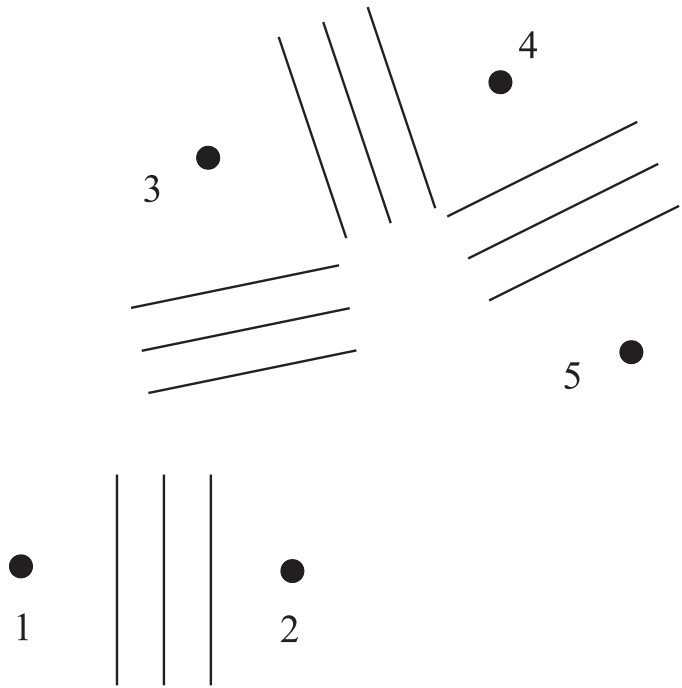}
\caption{The Feynman propagator can do without a manifold
diffeomorphism. We have a Feynman fabric of ${\Large \bullet }$
random culminating points and $\mid \mid \mid $ Dirac phase
factors between. This fabric is over the tangent objects of the
points. \label{fig3}}
\end{center}
\end{figure}

Remark to the tangent objects (appendix C). An electrodynamic (physical)
space structure is separately defined for each tangent object. The tangent
object is an ''electrodynamic neighborhood'' of each culminating point. A
common mathematical Minkowski structure for ''before'' and ''after'' a
particle collision, however, is not a prior of the hidden charge model.%
\newline
\newline

Each physical tangent object can methodically be stratified into layers
which contain different physical aspects of the potentials from the
different eigensolutions on the $H$ space. Examples. The first layer is an
algebraic construction of a biquaternion space \cite{Donth1988}, and the
second layer is specified by ''physical coordinates'' marked by the
potential components of the eigenfunctions for different classes. This
results in an electrodynamic Minkowski structure $M_{4}$ for the tangent
object for 1-class eigensolutions (e.g. leptons) and in a fourdimensional
Euclidean structure $E^{4}$ for 2-class eigensolutions (e.g. dark matter
particles or the prebaryon). The common space is a mix of tangent objects
with different specification, e.g. a mix of $M_{4}$ and $E^{4}$ tangent
objects (crumbled space, section 4.1). This is not a manifold.\newline
\newline

Remark to the path integrals (figure 3): particle collisions result in
Feynman diagrams from the elements of the Feynman fabric over the tangent
objects of the particles.\newline
\newline

A prior of the conventional gauge-theory approach to the Standard model ($%
SU(2)$, $SU(3)$, ..., GUT unification) is based on absolutizing the space
{\it metric}: the real space be a manifold with its diffeomorphism for the
metric, also for physics without gravitation. This diffeomorphism would
generate a topological space unifying points and filters as neighborhoods.
This would destroy the $x/\psi $ separation of figure 2.\newline
\newline

The $x/\psi $ separation maintains the chance of our model to be consistent
with the confirmed part of the Standard Model: 1). The Feynman propagator
(figure 3\cite{DonthWZ89}) can do without the diffeomorphism of a
mathematical space manifold. 2). Electrodynamics ($dF=0$, $d\,~^{\ast
}F=e~^{\ast }j$) and invariant vector fields of a group space need only
external differentials $d$ for our physical applications. 3). As mentioned
above, the conformal electromagnetic wave metric on $H$ (equation \ref%
{Eq.7.1}, below) has no intrinsic length, but only angles. 4). The quantum
mechanical relation (from the Feynman path integral) between mass $m_{0}$
(e.g. for an exchange boson) and length $l$, $m_{0}\sim 1/l$, can do with
the mass definition from the culminating point and a length $l$ mediated by
the tangent objects. This $m_{0}\sim 1/l$ relation does not need the
diffeomorphism.\newline
\newline

Once again in other words: The massless model photon alone cannot define a
metric tensor on the tangent object. Nevertheless, electromagnetic
properties are essentially used to construct the tangent object (as
discussed in appendix C). The mathematical diffeomorphism corresponds
physically to the motion of the model graviton. This graviton is
fundamentally connected to large space scales and is ''very weak'' (a shadow
with factor 10$^{-40}$) in comparison to the ''stronger'' electromagnetic
properties of our flat $M_{4}$ tangent object in the empty space.\newline
\newline
What things (besides a strong charge symmetry) can we obtain from the model?%
\newline
\newline

{\it Section 3}: Although there are no particles and no parameters in the
hidden charge, we get a simple and clear system of particles in the reality
space from the only three classes of eigensolutions.\newline
\newline

{\it Section 4}: The conventional electroweak Lagrangian in $M$ is obtained
from the $H\mapsto M$ map of the invariant vector fields of $H$. Instead of
Higgs (no relevant eigensolutions for Higgs particles and no energetic cause
for mass), a mass operator with action on the $\vartheta $ segment for the
filter is detected. Quantum field theory is explicitely obtained via Feynman
path integrals; quantum theory seems possible also for 2-class particles
with $E^{4}$ tangent objects, if they are isolated and imbedded in only $%
M_{4}$ tangent objects of the crumbled space.\newline
\newline

{\it Section 5}: From the filter construction we get numerical values for
the Weinberg angle $\theta _{W}$, the current coupling constant $\alpha
(\sin ^{2}\theta _{W})$, and the mass curve of charges leptons ($e,\mu ,\tau
$) with an existential instability for higher families \cite{Donth1988}.%
\newline
\newline

\subsection*{2.3 Wherefrom do the many elements for the filters come?}

The existential instability in the whole model comes from the maximum of the
mass curve of charged leptons (figure 6, below, cf. also figure 2 in \cite%
{Donth1988}). Existential stability includes that the mass of a
corresponding particle in the next family is larger. An instability would
enter, if the mass in the next but one and the following families become
smaller. Then, a particle in the next family cannot exist. The boundary
defines the number of existing charged lepton families; e.g. $\tau =3$ would
be the last existing, if eka$~\tau =4$ would have a larger mass, but eka$%
~\tau =5,6,7,...\rightarrow \infty $ successively smaller ones. Existential
instability is sharper than the \ quantum mechanical one: Such particles
cannot exist even as virtual particles in Feynman diagrams. Example: The
tauon $\tau =3$ is quantum-mechanical instable but existentially stable.%
\newline
\newline

The mass operator contains only $\partial /\partial \vartheta $ (besides a
Hopf fiber for imbedding the particle point in its tangent object). The
filter concerns, therefore, only the $\vartheta $ segment coordinate. This
coordinate is contained in the harmonic potentials by the ''harmonic
polynomials'' $y_{mk}(z)$, $z=\cos ^{2}\vartheta $, with $m$ for charge and $%
k$ the family number (section 3.2). For the vacuum ($k\rightarrow \infty $, $%
m\approx 0$) the number of zeros $=k$, i.e. the size of the segment part is
of order $0(1/k)\rightarrow 0$. It is suggested that the convergence of
segment parts mediates some kind of a Cauchy filter with numbers for $z$
vacuum elements. From the polynomials we get huge numbers (section 3.2) that
facilitate the handling of the filter convergence for coupling constant and
masses.\newline
\newline

How get the other particles their masses? If they (as common particles from
several eigensolutions, e.g. hadrons, or with $E^{4}$ tangent objects, e.g.
dark matter) are imbedded in $M_{4}$ tangent objects of the crumbled space,
I expect, seen from outside, similar methods as for charged leptons and,
therefore, similar limitations from existential stability. {\it Example}:
Since the $z$ structure of the dark matter particle is similar to the $z$
structure for the model electroweak exchange bosons, a dark matter particle
gets an estimated mass of order 100 GeV. The $z$ structures are: for (1,1)
dark matter $\sqrt{z/(1-z)}$\cite{Donth2006}, for $W^{\pm }$ exchange bosons
($\sqrt{z},\sqrt{1-z}$), and for the $Z$ exchange boson $\sqrt{z(1-z)}$. The
two latter structures lead to $m_{0}(W^{\pm })/m_{0}(Z)=\cos \theta _{W}$ as
expected for the mass ratio of the exchange bosons.

\subsection*{2.4 Hidden charge model vs pristine string model}

String particles do not have physical elementary as defined in
section 1. String particles are excitations of (in the pristine
form: onedimensional) objects consisting of ''many'' points of a
manifold; the string particles are much larger than the points. In
the hidden charge model, however, the particles are elementary
points in the reality space, generated by the converging filters
of culminating points. The space points have physical elementarity
in the above sense by being possible locations of elementary
particles.\newline
\newline

In addition, in the string theory, space-time and physics are separated in
conventional manner. The prior of a higher-dimensional Riemann-Klein-Kaluza
space-time manifold consistent with physics carries a physical Lagrangian
and a physical Higgs mechanism. Complications by e.g. warping and sequesting
the space are accepted in the string community. In the hidden charge model,
the borderline between space-time and physics is complicated from the
beginning. E.g., the tangent object is not only a geometrical object but is
physically motivated and stratified in different layers (see table 2 in
appendix C of section 7, below).\newline
\newline

Moreover, the number of dimensions for string theory is 10 or 11, i.e. 6 or
7 of them are compactified, i.e. hidden at small lengths corresponding, in
the pristine versions, to high (Planck) energy: 10$^{-20}$ fermi and 10$%
^{19} $ GeV, respectively. The hidden charge model remains in four
dimensions. The hidden $H$ space is a compact $S^{1}\times S^{3}$ manifold \
with conformal invariance. This ensures omnipresence and application (the
map of figure 2) to e.g. the very large universe as well as to Compton wave
lengths of hadrons. The map between $H$ space and reality $M$ space is
initiated by existential instability for higher particle families.\newline
\newline

Additionally, the number of eigensolutions in the string theory is usually
very large, and the separation of an experimental range (%
\mbox{$<$}
1000 GeV, say) is a thankless task. In the hidden charge model, the number
of particles is restricted by only three classes of eigensolutions, and to
low energies due to existential instability, so that a clear picture can be
developed from the very beginning.\newline
\newline

String theory delegates crumbling of space to the small Planck length scale.
The hidden charge model sees a crumbled space from the $x/\psi $ separation
as a general companion for quantum theory in all length scales. Conventional
quantum theory (so far successful for electroweak interaction) is, as seen
from our model, a trial to define the circumstances on the basis of a
manifold.\newline
\newline

String theory includes not only a unification of electroweak and strong
theory at the GUT scale ($3\cdot 10^{16}$ GeV), but also the gravitation can
be included. This suggests a ''hot'' unification at the above Planck scale
(10$^{19}$ GeV) and a hot start of the universe. The hidden charge model,
however, allows a cold unification ($\lesssim 10^{3}$ GeV) and a cold start,
mainly because of the unusual properties of our dark matter particles \cite%
{second part} and because a metric diffeomorphism $g(x)$ is not assumed as a
prior.\newline

\section*{3. Stock of particles from Maxwell-type eigensolutions for $A$
potentials on the hidden $S^{1}\times S^{3}$ charge}

The hidden $H$ space suits well to electrodynamics. It has four dimensions
and allows a Minkowski (wave) metric in angles (though with no $c$ constant
in meters/second for conformal invariance of wave fields). Biharmonic
coordinates (appendix A in section 7, below): $S^{1}$ with $0\leq \tau \leq
2\pi $, $S^{3}$ with $0\leq \varphi _{1},\varphi _{2}\leq 2\pi $, $0\leq
\vartheta \leq \pi /2$, allow a simple definition of a $\oplus
\leftrightarrow \ominus $ charge symmetry $C$ on the basis of only wave
fields (div$_{4}$ $A=0$), reflecting the equivalence of the two Heegaard
tori in $S^{3}$ (figure 7 in appendix B of section 7, below):

\begin{equation}
C:(\oplus ,\tau ,(\varphi _{1},m_{1}),(\varphi _{2},m_{2}),z=\cos
^{2}\vartheta )\leftrightarrow (\ominus ,\tau ,(\varphi
_{2},m_{2}),(\varphi _{1},m_{1}),\zeta =\sin ^{2}\vartheta ).
\label{Eq.3.1}
\end{equation}%
\newline

The three onedimensional tori 0...2$\pi $ come \cite{Busse1998} from the
Cartan subalgebra of $u(2,2)$. The latter is the algebra for the conformal
group Conf$(M_{4})$ with 15 parameters; the Lie algebra $u(2,2)$ is
equivalent to $so(4,2)$. A Cartan subalgebra is the maximal commutative
subalgebra and has here three dimensions: Maxwell's: equations in $A$ on $%
S^{1}\times S^{3}$ (appendix B in section 7, below) have three such toris
with corresponding torus factors in the potentials,

\begin{eqnarray}
\partial _{\tau }A_{\tau } &=&i\omega A_{\tau }\text{ with }e^{i\omega \tau
}, \label{Eq.3.2} \\
\partial _{\varphi _{1}}A_{1} &=&im_{1}A_{1}\text{ with }e^{im_{1}\varphi
_{1}},\text{ and }\partial _{\varphi _{2}}=im_{2}A_{2}\text{ with }%
e^{im_{2}\varphi _{2}}.  \nonumber
\end{eqnarray}%
\newline

These tori are generally connected with the signatures of wave functions $%
\psi $ on the tangent objects in the reality $M$ space. The $\vartheta $
segment is separated by ($A_{\vartheta }\equiv 0$, no torus) and is used for
culminating points $x$. The potential $A$ instead of the Maxwell field
tensor $F$, $F\sim dA$, was chosen because of its occurrence in the Dirac
Phase factors (Aharonov Bohm effect of quantum mechanics).

\subsection*{3.1 Three classes of eigensolutions}

A complete classification \cite{Busse1998} can then be based on properties
of the $S^{1}$ torus component $A_{0}=A_{\tau }$; there are three classes
(table 1) whose mathematical details for 1-class are in \cite{DonthLange}
(harmonics with harmonic polynomials), and for 2-class and 3-class in \cite%
{Busse1998}. The verbal terms for the \cite{Busse1998} model use are
partially redefined after the cosmological revolution \cite{Donth2006}.

\begin{table}[tbp] \centering%
\caption{$S^1$ Classification of eigensolutions [16,17]\label{key}}%
\begin{tabular}{l|l|l|l|l|l|}
class & potential & tangent & typical & specification & model use \\
& components & objects & functions &  &  \\
& (physical & (wave &  &  &  \\
& coordinates) & equation) &  &  &  \\ \hline
1-class. & $A_{1}(\varphi _{2},\vartheta ,\tau )=$ & $M_{4}$ & harmonic &
neutral: $m_{2}=0$ & neutrinos. \\
&  &  & polynomials & $\tau $ torus &  \\ \cline{5-6}
$\frac{\partial A_{0}}{\partial t}=0$, & $=e^{i\omega \tau -im_{2}\varphi
_{2}}\cdot y_{mk},$ & (hyperbolic) & (equation (3.3)), &  &  \\
$A_{0}\equiv 0$ & $\omega =2k+m.$ &  & $y_{mk}(z),$ & charged ($m_{2}\neq 0$)
& ($e,\mu ,\tau $) \\
& $(\varphi _{1})$\qquad FN$^{\text{a}}$ &  & $z=$cos$^{2}\vartheta .$ & $%
\tau ,\varphi _{2}$ tori & charged \\
&  &  &  &  & leptons. \\ \hline
2-class. & $A_{0}=e^{i(m_{1}\varphi _{1}+m_{2}\varphi _{2})}\times $ & $%
E^{4} $ & hypergeometric & one-charge & prebaryon. \\
$\frac{\partial A_{0}}{\partial \tau }=0$, & \qquad $\times A_{0}(\vartheta
).$ & (elliptic) & $A_{0}(\vartheta )$ & $\varphi _{2}$ tori & (baryon \\
$A_{0}\neq 0$ & ($\tau $) &  & FN$^{\text{b}}$ & ($m_{2}\neq 0$), & charge
\\
&  &  &  &  & $m_{2}=1$) \\ \cline{5-6}
&  &  &  & or $m_{2}>m_{1}>0$ &  \\ \cline{5-6}
&  &  &  & neutral two- & dark matter \\
&  &  &  & charge tori &  \\
&  &  &  & ($\varphi _{1},\varphi _{2}$) &  \\
&  &  &  & $\left| m_{1}\right| =\left| m_{2}\right| >0$ & FN$^{\text{c}}$
\\ \hline
3-class. & $A=(A_{0},A_{1},A_{2},0)$ & $M_{4}$ & the one special & all $%
\left| \text{charges}\right| $ & confinons \\
$\frac{\partial A_{0}}{\partial \tau }\neq 0$ & $=e^{i(4/3)(\sigma _{0}\tau
+\sigma _{1}\varphi _{1}+\sigma _{2}\varphi _{2})}\times $ & (hyperbolic) &
solution with 8 & equal to 4/3 & for \\
& $\times (\sin \vartheta \cos \vartheta )^{-4/3}\times $ &  & possibilities
for & ($\omega =4/3,$ & hadrons. \\
& $\times (\sigma _{0},\sigma _{1}\cos ^{4}\vartheta ,\sigma _{2}\sin
^{4}\vartheta ,0)$. &  & $\sigma _{0},\sigma _{1},\sigma _{2}\in $ & $%
m_{1}=4/3,$ &  \\
& $(\tau ,\varphi _{1},\varphi _{2})$ &  & $\in (+1,-1)$ & $m_{2}=4/3$). & FN%
$^{\text{d}}$%
\end{tabular}%

Footnotes.

FN$^{\text{a}}$: Correspondingly for $A_{2}$ from charge symmetry
$C$, equation (3.1). The polynomials are in equation (2.3), and
the vacuum
elements in section 3.2. FN$^{\text{b}}$: Examples for the amplitudes $%
A_{0}(\vartheta )$ are the prebaryon, $z^{1/2}F(\frac{1}{2},\frac{3}{2},2,z)$%
, dark matter (1,1), $(z/(1-z))^{1/2}$, ($m,m$), $(z/(1-z))^{m/2}$ for $%
\oplus $ particles. FN$^{\text{c}}$. 2-class particles can be
measured as usual in the isolated pure state (e.g. the mass of a
dark matter particle) or in a caged state (e.g. the energy
contributions to the baryon from the prebaryon), if the outside is
the $M_{4}$ tangent object, e.g. in the ''vacuum''.
FN$^{\text{d}}$. One hadron consists of two (for mesons) or three
(for baryons) leptons captured by one confinon; the captured
leptons are then called quarks. For baryons, additionally one
prebaryon is captured introducing the baryon charge (section 5.4).

\end{table}%

The eigensolutions suggest leptons, dark matter particles, prebaryons,
baryon charges, and confinons used for hadronic confinement (section 5.4),
with a model approach \cite{Donth2006} to strong interaction instead of the
conventional $SU(3)$ gauge approach. The classification of table 1 excludes:
Any independent, fluctuating (and therefore quantizable) vacuum field (i.e.
no Higgs field and no Higgs particle in the model, cf. section 3.2); any
supersymmetry (details in appendix C, section 7, tangent object
construction), and isolated quarks with third charges (the model quarks are
confined leptons, the third charges come from the confining confinon), there
is also no quark gluon plasma at high temperatures in the model. The mass
scale is limited by the order 1 TeV = 1000 GeV, if the existential
instability is transmittable to all kinds of isolated particles; we have
vacuum elements suitable for filter constructions only from 1-class vacuum
solutions. The culminating point / tangent object $\ \ x/\psi $ separation
is maintained for all isolated particles.\newline
\newline

\subsection*{3.2 Vacuum elements. No Higgs}

Be $k$ the family number, $\kappa $, $0\leq \kappa \leq k$, the dummy
integer, and $m\geq 0$ the integer charge number for 1-class solutions. Then
for the potential $A_{1}$ (analogously for $A_{2}$ of the antiparticle from
charge symmetry $C$, equation (\ref{Eq.3.1})) we get the polynomial for the
1-class solution \cite{Donth1986,DonthLange}

\begin{equation}
y_{mk}(z)=z^{m/2}%
\mathop{\textstyle\sum}%
\limits_{\kappa =0}^{k}(-1)^{\kappa }%
{k \choose \kappa }%
{k+\kappa +m-1 \choose k-1}%
z^{\kappa }; \label{Eq.3.3}
\end{equation}

\begin{equation}
A_{1}^{0k}(z)=\exp (2k\tau )y_{ok}(z), \label{Eq.3.4}
\end{equation}

\begin{equation}
y_{ok}(z)=%
\mathop{\textstyle\sum}%
\limits_{\kappa }(-1)%
{k \choose \kappa }%
{k+\kappa -1 \choose k-1}%
z^{\kappa } \label{Eq.3.5}
\end{equation}%
with $z=\cos ^{2}\vartheta $. The $m=0$ solutions, $A_{1}^{0k}(z)$ with $%
y_{0k}(z)$, are for neutrinos. Formally, a vacuum solution ($m=0$ or $m\ll k$%
, $k\rightarrow \infty $) may be called ''$\infty $ neutrino''.\newline
\newline

If, however, existential instability can also be established for the
neutrinos (as discussed in section 5.3), then large numbers of $z$ in $%
(z)^{\kappa }$, called {\it vacuum elements}, are always obtained,
especially when the decay goes down to the single $z=(z)^{1}$ (this is
called {\it electronic vacuum} in the filter). The number of vacuum elements
is then a

\begin{equation}
\text{huge number }=%
{k \choose \kappa }%
{k+\kappa -1 \choose k-1}%
. \label{Eq.3.6}
\end{equation}%
\newline

To get some kind of a Cauchy filter, we assume that a ''distance'' $d$ can
be defined by the $\vartheta $ angle distance between the $y_{mk}(z)$ zeros
in the $\vartheta $ segment. As then $d=0(1/k)$, ''smaller'' vacuum elements
are obtained for larger $k$, their number hugely increases with $k$. If,
therefore, the hidden charge with the conformal wave metric (7.1)-(7.3) is
taken as a primary basis for the filter, then the filter is a (Fr\'{e}chet)
Cauchy filter (section 2.1): Since the filter is a part of the $H\mapsto M$
map, the vacuum elements contain, as image constraints, also some
information about lengths in the $M$ space. This leads to the concept of
filter elements, explained in \cite{second part} (at the beginning of
section 5.1 there).\newline
\newline

The vacuum elements are no eigensolutions, because an individual torus $%
S_{\tau }^{1}$ is missing for them: $A_{1}^{0k}\neq (A_{1}^{01})^{k}$ for $%
k>1$. This means no independent particle can be connected with vacuum
elements alone. The huge numbers (\ref{Eq.3.6}) prevent any kind of
fluctuation, i.e. there is no independent and general vacuum field that
could be quantizable. Physically, the vacuum elements form the
(non-quantizable) filter of the mass point; neither Higgs fields nor Higgs
particles are constituents of the model.

\section*{4. Electroweak Lagrangian in physical space from invariant vector
fields in the hidden space. The mass operator}

The physical and mathematical roots in the ''layers'' of tangent objects are
listed in the appendices C, D, and E: In C the construction of a
biquaternion $M$ space for each tangent object from wave metric and charge
symmetry of the $H$ space, the way to $E^{4}$ or $M_{4}$ tangent objects
from physical coordinates, and the construction of exchange bosons; in D the
$u(2)$ invariant vector fields; resulting in the empirical (Weinberg Salam)
electroweak sector of the Lagrangian recapitulated in E.\newline
\newline

The main result is obtained from a term-by-term comparison of the invariant
vector fields with the empirical (conventional) Lagrangian in the
corresponding exponentials. No rests are left in D and E, i.e. neither in
the fields nor in the Lagrangian. The Lagrangian is obtained from the fields
if (1) The Weinberg angle $\theta _{W}$ is a definite value ($\overline{%
\vartheta }$) of the $\vartheta $ segment coordinate used for the filter.
(2) The mass operator is a differentiation in the direction of the filter
coordinate $\vartheta $ and substitutes the Higgs construction of the
conventional Lagrangian:

\begin{equation}
\text{Weinberg angle}:\text{ }\theta _{W}=\overline{\vartheta
},\qquad \qquad \qquad \qquad \qquad \quad \text{\ }
\label{Eq.4.1}
\end{equation}

\begin{equation}
\text{Mass operator}:\text{ }\partial =%
{\sin \varphi  \choose \cos \varphi }%
\partial _{3}\text{ with }\partial _{3}=\partial /\partial \vartheta ,
 \label{Eq.4.2}
\end{equation}%
where $\varphi =\varphi _{1}+\varphi _{2}$ is the fiber of the chosen Hopf
bundle card $S^{3}\rightarrow S^{2}$.\newline
\newline

A full proof was announced in 1991 \cite{DonthWZ89,DonthWZ265} and will be
published in \cite{Donth2006}. The matter is \ complicated from two aspects.

\begin{enumerate}
\item The exchange bosons are binary torus constructions (table 2 and 3 in
appendix C): We need rules for torus handling, e.g. which operators of the
invariant vector fields activate the tori ({\it torus induction}, table 3).

\item {\it Image constraints}: In the comparison, seen as a map, the primary
set of $M$ space elements (points, connections, tangent objects) is the
origin (figure 2) and the conventional Minkowski $M_{4}$ manifold used for
the formulation of the conventional Lagrangian is (part of) the image. If
the primary set is finer than (and not so ''dense'' as) the conventional $%
M_{4}$ manifold, then the comparison is constrained by the harder properties
of the coarse tight manifold used for the conventional Lagrangian (section
4.1, below).\newline
\newline
\end{enumerate}

The results (\ref{Eq.4.1}-\ref{Eq.4.2}) confirm two things of our model: A
clear distinction of mass and energy, and a quantum theory from our
''classical'' hidden charge. The mass generation is ascribed to the $%
\vartheta $ coordinate of the filter construction for the culminating point.
The energy, on the other hand, is a construction from the wave functions $%
\psi $ of the Feynman fabric (figure 3) on the tangent objects. The wave
equation for $\psi $ contains the mass values of the point particles, the
Weinberg angle, the coupling parameters and (all the) other parameters from
the filter, having in this way influence on the energies, when the quantum
mechanical wave equation is solved. [According to the model, the field $\phi
$ of appendix E (for the conventional Yukawa mass terms) is not an
independent Higgs field, but at the best some symbolic trial to connect the
filter with a field.]\newline
\newline

The quantum theory of the model starts from the charge with their tori, and
not from the Planck constant $\hbar $. The partition into the above primary
set of space elements and their bond by the hidden charge evaluates the
model to be some realization of Pauli's vision of such a quantum field
theory which also determines its numerical parameters. \ (Let us repeat,
that there are no parameters in the conformal hidden charge, even no speed
of light in meters/second). The probability aspects are introduced by the
randomness of tori coordinates ($\tau ,\varphi _{1},\varphi _{2}$) where the
$\vartheta $ segment for the filter is pinned at them. After the map, the
randomness is regulated by the Lagrangian from the invariant vector fields
determining the probability density $\left| \psi (r)\right| ^{2}$ and the
transition probabilities. The electroweak Lagrangian is not put from gauge
principles, but is derived solely from the $u(2)$ symmetry of the hidden $H$
space. This leads directly to the Dirac phase factors for finite
''rotations'', introducing $\hbar $ and the quantum field theory from
Feynman path integrals. The generality of quantum theory is restricted to a
Minkowski $M_{4}$ environment and is more detailedly discussed near the end
of \cite{second part}).\newline

\subsection*{4.1 Examples for image constraints. Crumbled space}

The mix of tangent objects with different specifications (e.g. spin
directions, $E^{4}/M_{4}$ fractions) is called a {\it crumbled space}. This
space has no metric diffeomorphism. Its map into {\it conventional} physical
manifolds (e.g. $M_{4}$ for electroweak interactions, $g(x)$ for general
relativity \cite{second part}) is characterized by image constraints,
because the crumbed space is ''finer'', not so restricted than the
''coarse'' manifold:

\begin{equation}
\text{fine origin (crumbled space) }\longmapsto \text{ coarse
image (manifold).} \label{Eq.4.3}
\end{equation}%
Examples: (1) The crumbled space allows two $\vartheta $ directions (figure
4, section 5.1 below), but the $M_{4}$ manifold must do with one Weinberg
angle $\theta _{W}$. For the neutral current $e_{L}$ (7.28) e.g. we must use
the internal $\vartheta $ orientation and obtain $\tan \theta _{W}-\cot
\theta _{W}$, for the neutral current $e_{R}$ (7.31), however, we must use
the external $\vartheta $ orientation and obtain $\tan \theta _{W}+\tan
\theta _{W}$ from the operator $\cot \vartheta \cdot \partial _{1}-\tan
\vartheta _{2}$ of the vector fields, because in the standard model $e_{L}$
is only one component of the pair $%
{\nu  \choose e_{L}}%
$, and $e_{R}$ is alone \cite{Donth2006}. (2) Pauli's spin statistics
theorem follows from a ''projection'' aspect of the map (\ref{Eq.4.3}),
since, for each pair of spins with opposite spins, a common 180$%
{{}^\circ}%
$rotation of the pair is equivalent to the exchange of the partners \cite%
{Donth1970}. Crumbling means here that the pairs have different common spin
directions. This construction excludes any SUSY construction from the
particle list of our model (table 1, also not containing any SUSY partners).
(3) A nontrivial mixture of $E^{4}$ (e.g. dark matter) / $M_{4}$ (others)
tangent objects is used for an initial liquid before and during the
cosmological inflation. This allows a cold big bang ($\lesssim $ 1000 GeV,
section 5.2 of \cite{second part}).\newline
\newline

The image constraints are an important methodological tool in our model.%
\newline

\section*{5. Calculation / Estimation of parameters}

This section is to calculate or estimate the Weinberg angle, $\theta _{W}$,
the current coupling constant of the electroweak sector, $\alpha (\sin
^{2}\theta _{W})$, and \ the mass values of the charged leptons ($%
m_{0e},m_{0\mu },m_{0\tau }$). The estimation becomes more complicated in
this succession, because e.g. the mass estimation supposes the action of the
mass operator (\ref{Eq.4.2}) in the filter. We use some kind of a {\it %
huge-number approximation}, that the estimation can be based on the
increasingly huge numbers of vacuum elements (\ref{Eq.3.5}) alone. This
means that the amplitudes, the details of segment partition etc. will be
neglected. We must expect that the precision of estimations decreases in the
above succession.\newline

\subsection*{5.1 Weinberg angle}

The Weinberg angle of the low-energy limit can be estimated
\cite{DonthWZ89} from the consistency of an external and an
internal view of the image constraint for $\vartheta \mapsto
\theta _{W}$ (figure 4a and 4b).

\begin{figure}[here]
\begin{center}
\includegraphics[width=0.5\linewidth]{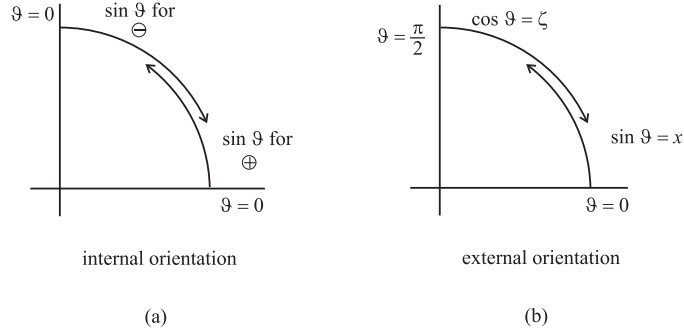}
\caption{Two orientations of the $\vartheta $ segment coordinate.
\label{fig4}}
\end{center}
\end{figure}

The internal orientation is based on the strong charge symmetry $C$ (\ref%
{Eq.3.1}) of the hidden space. The coordinate values of the $\vartheta $
segments must be the same for $\oplus $ and $\ominus $ charges. The internal
measure for $\vartheta $ is defined by the Haar measure for the $u(2)$ group
algebra, $d\sqrt{-g}\sim d\sin ^{2}\vartheta $ (appendix A, wave metric
equation (\ref{Eq.7.3})). The external orientation distinguishes the $\oplus
$ and $\ominus $ charges; the two corresponding cards of the $%
S^{3}\rightarrow S^{2}$ Hopf map with corresponding fibers $\varphi _{1}\pm
\varphi _{2}$ have the external measure $d\vartheta $. The low-energy limit
allows to consider only the lowest vacuum elements: $z=\zeta ^{2}=\cos
^{2}\vartheta $ or $x=\xi ^{2}=\sin ^{2}\vartheta $, for the two cards.
Defining a mean value of a function $\widetilde{f}(z)$ by

\begin{equation}
\left\langle \widetilde{f}(z)\right\rangle ^{\pm }=\frac{2}{\pi }%
\textstyle\int%
\limits_{0}^{\pi /2}d\vartheta ~f\,(\vartheta ^{\pm })\quad ,\quad
\vartheta ^{\pm }=\vartheta \pm \vartheta _{0}, \label{Eq.5.1a}
\end{equation}%
we get from $\widetilde{f}(z)\sim z$ (or $x$) \ for the low-energy limit of $%
y_{0}=\sin ^{2}\vartheta _{0}$ the equation: ''internal'' = ''external'' %
\cite{Donth1986,DonthWZ89}

\begin{equation}
y_{0}=\frac{1}{2}-\frac{2}{\pi }[y_{0}(1-y_{0})]^{1/2},
\label{Eq.5.1b}
\end{equation}%
with the solution

\begin{equation}
y_{0}=\frac{\sqrt{\pi ^{2}+4}-2}{2\sqrt{\pi ^{2}+4}}=0.231\,458\,363\,927...%
\text{ .} \label{Eq.5.2}
\end{equation}%
The approximate result 0.23146 is in a good correspondence to the
experimental values of the effective Weinberg angle 0.23149(13) \cite%
{particle 4a}. The value $\sin ^{2}\vartheta ^{\prime }\approx 0.2315$ was
quoted in 1986 \cite{DonthLange}. The $y_{0}$ value is a general parameter
of the electroweak sector of our model and does not depend on the family
number.\newline

\subsection*{5.2 Current coupling constant}

We try a connection between Feynman diagram renormalization with the
huge-number approximation (\ref{Eq.3.5}) of the filter renormalization: $%
\kappa \rightarrow \infty $, $k\rightarrow \infty $, $y=\kappa /k=$ finite.
The order factor of the diagram, $\alpha ^{\kappa }$, is substituted by the
filter prefactor

\begin{equation}
\alpha ^{\kappa }\rightarrow g^{\kappa }%
{k \choose \kappa }%
{k+\kappa -1 \choose k-1}%
=\text{ finite.} \label{Eq.5.3}
\end{equation}%
Finiteness is required for any kind of experiment for a general observable $%
X $. From equation (\ref{Eq.5.3}) we obtain a finite current coupling
constant, $\alpha (y)$, from the finiteness of the experimental result $%
\left\langle X\right\rangle $,

\begin{equation}
\left\langle X\right\rangle \propto g^{\kappa }%
{k \choose \kappa }%
{k+\kappa -1 \choose k-1}%
=\left\{
\begin{array}{ccc}
0 & \text{for} & g<\alpha (y) \\
\text{finite} & \text{for} & g=\alpha (y) \\
\infty & \text{for} & g>\alpha (y)%
\end{array}%
\right\} \label{Eq.5.4}
\end{equation}%
and put $y=\sin ^{2}\theta _{W}$ as the simplest internal measure. From (\ref%
{Eq.5.4}) we calculate for huge numbers

\begin{equation}
\alpha (y)=y^{2}(1-y)^{\frac{1-y}{y}}(1+y)^{-\frac{1+y}{y}}
\label{Eq.5.5}
\end{equation}%
which function is graphed in figure 5.

\begin{figure}[here]
\begin{center}
\includegraphics[width=0.5\linewidth]{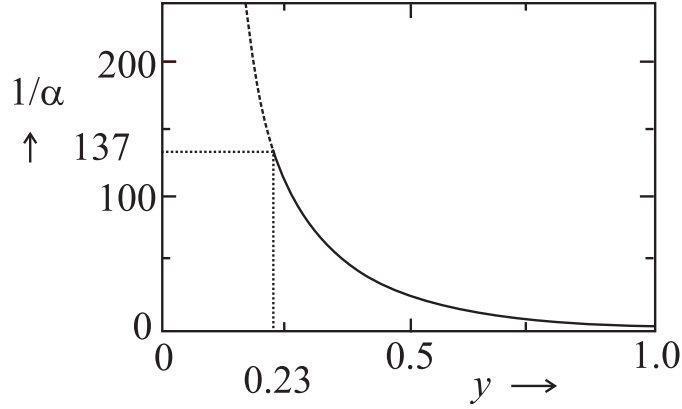}
\caption{Reciprocal current coupling constant $1/\alpha $ as
function of a current Weinberg angle $y=\sin ^{2}\theta _{W}$ from
the renormalization in the huge-number approximation
(\ref{Eq.5.4}) for the filter. \label{fig5}}
\end{center}
\end{figure}

Special values from the approximation are the low energy limit $\alpha
^{-1}(y_{0})=135.411~830~525$ for $y_{0}$ of equation (\ref{Eq.5.2}), $%
\alpha ^{-1}(1)=4$, $\alpha ^{-1}(1/2)=27$, and $\alpha ^{-1}(3/8)\approx
50.0319$.

\subsection*{5.3 Mass of charged leptons}

To use the huge-number approximation similar to equation (\ref{Eq.5.4}), the
$z$ components for the charged leptons must now be taken into consideration
(equation (\ref{Eq.3.3}) with charge $m=1$ for low families $k_{0}=1,2,3,...$%
). We assume that mass renormalization via the filter can be described by
numbers for the possibilities of mistaking; also the $z$ values of the
individual leptons with the huge numbers (\ref{Eq.3.6}) of general vacuum
elements $z$ in (\ref{Eq.3.3}). There are three possibilities

\begin{equation}
\left.
\begin{array}{ccc}
\text{I individual - individual\quad :} & \text{particle factor} & \Pi
(k_{0})\text{ , }k_{0}=1,2,3,... \\
\text{II general - general\quad\ \ \ \ \ \ :} & \text{vacuum factor} &
P_{V}\,(\kappa ,k) \\
\text{III individual - general\quad :} & \text{particle-vacuum factor} &
P_{L}(k_{0},\kappa ,k)%
\end{array}%
\right\} \label{Eq.5.6}
\end{equation}%
The factor I is a damping factor, and II and III are amplication factors for
mass. Putting the Compton wave length $\lambda _{0}\sim 1/m_{0}$, with $%
m_{0} $ the mass value, we have now

\begin{equation}
\lbrack \lambda _{0}\Pi (k_{0})]^{\kappa }\cdot P_{V}(\kappa
,k)\cdot P_{L}(k_{0},\kappa ,k)=\text{ finite.} \label{Eq.5.7}
\end{equation}%
The analogy to equation (\ref{Eq.5.4}) assures the scaling of the mass
spectrum by the coupling constant $\alpha $.\newline
\newline

The result is a maximum of the mass curve $m_{0}(k_{0})$, because the
damping factor varies stronger than the amplication factors, approximately

\begin{equation}
\exp \Pi \sim k_{0}\ln k_{0}\text{ , }P_{V}\cdot P_{L}\sim \text{
sums of polynomials.} \label{Eq.5.8}
\end{equation}%
The maximum was firstly published in 1988 \cite{Donth1988} and
supports the concept of existential instability of section 2.3
(figure 6).

\begin{figure}[here]
\begin{center}
\includegraphics[width=0.5\linewidth]{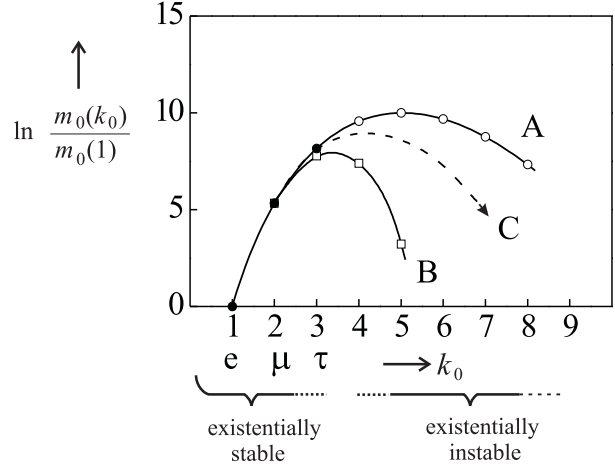}
\caption{Estimated mass curve for charged leptons. A.
''Thermodynamic variant'' equation (\ref{Eq.5.9}), adjusted to the
two known experimental ratios for $\mu /e$ and $\tau /e$. B.
''Statistical variant'' of the text. C. Ad hoc example that would
give the existential stability of the three observed families.
\label{fig6}}
\end{center}
\end{figure}

In a ''thermodynamic variant'', equation (\ref{Eq.5.8}) is used for the
estimation,

\begin{equation}
\ln m_{0}=G+k_{0}F-k_{0}\ln k_{0}U, \label{Eq.5.9}
\end{equation}%
where the ''potentials'' $G$, $F$ and $U$ are assumed to depend slowly on
the family number $k_{0}$: We put them constants. After adjustment at the
known experimental values the curve A is obtained, allowing four
existentially stable families.\newline
\newline

In a ''statistical variant'' with a mass operator differentiation $\partial
/\partial \vartheta $ in one component of a bilinear product of the
polynomials, with an optimization ($dQ/du=0$) for $u$ in the common
amplification factor, $P=P_{V}\cdot P_{L}$, assumed to be $%
[P(z,k_{0})]^{\kappa }=z^{\kappa u}\exp [\kappa Q(u)]$, and using $\alpha
\cdot z=1$, the curve B is obtained, allowing only two existentially stable
families. The start value of the mass curve, however, is obtained without
adjustment as $\ln (m_{0\mu }/m_{0e})=5.34$, near the experimental value
5.33.\newline
\newline

A mass curve with three families seems possible after refinement of the
methods (ad hoc hand waving curve C in figure 6).\newline
\newline

In the huge-number approximation, the neutrinos get zero masses, because no
Hopf fiber $\varphi $ (or $\psi =\varphi _{1}-\varphi _{2}$) can directly be
constructed for their mass operator. The polynomials for a damping factor,
however, would exist, say, for indirect higher-order approximations with
borrowed fibers from other collision partners. $-$ No Higgs fields or Higgs
particles are necessary for our general mass-curve estimations, based only
on the decay into 1-class vacuum elements (section 3.2) due to existential
instability.

\subsection*{5.4 Outlook for strong interaction}

The standard model of elementary particles and our hidden charge model have
different, alternative principles for the construction of a strong
interaction: The former is based on a special extension of the gauge theory
beyond the electroweak group: $U_{Y}(1)\times SU_{L}(2)\times SU_{C}(3)$,
the latter on the confinement into the frame of the three classes of
particles from table 1.\newline
\newline

Let us assume, that more complicated particles can be constructed by filters
for a common particle from eigensolutions of different classes.\newline
\newline

The general construction for hadrons is described in the footnote FN$^{\text{%
d}}$ of table 1 in section 3.1. For baryons, the following ''chemical''
lepton capture reaction can be considered:

\begin{equation}
\begin{array}{c}
\text{1-class leptons }\psi \text{ }+\text{ 2-class prebaryon }\chi ^{\prime
}+\text{ 3-class shell confinon }\varphi _{3}\rightleftharpoons \\
\rightleftharpoons \text{ quarks }q\text{ in confinement }+\Delta U\qquad
\qquad \qquad%
\end{array}
 \label{Eq.5.10}
\end{equation}%
\newline

The l.h.s. of the reaction is for the high temperature state with 1-class
leptons $\psi $, a single-charged 2-class prebaryon $\chi ^{\prime }$, and a
3-class confinon $\varphi _{3}$ (using a spatial interpretation as some
''shell'' cage from its three tori). The r.h.s. of the reaction is for low
temperatures: a baryon (or meson) from quarks captured inside of the cage
from the shell plus some binding energy $\Delta U$ partly supported, for
baryons, by the baryon charge from the prebaryon. Energy has its part in the
baryon mass: In the low temperature state we get an additional elliptic
binding potential from the 2-class $E^{4}$ tangent object of the prebaryon
in the crumbled space inside. The quarks have a limited independence, and
the gluons may be some kind of exchange bosons inside the shell.\newline
\newline

The conventionally expected phase transition is substituted by states with
different reaction partners after a shift of temperature (or pressure, ...).
We do not obtain a quark gluon plasma at high temperatures, but a mixture of
leptons, prebaryons, and confinons. With the latter we find always
three-piece animals, even at the high temperature state; further only the
ordinary leptons, charged or not (no quarks), and the prebaryon with $E^{4}$
tangent objects in the crumbled space, isolated or not. The existential
instability of higher leptons ensures an upper mass limit of particles with
strong interaction. The hidden charge model delivers numbers that are well
known from the $SU(3)$ gauge field construction: three physical coordinates,
three tori, neutralized third charges, and eight $\sigma $ combination
possibilities for the confinon of table 1.

\section*{6. Consequences for particle astrophysics}

I. The quaternion structure of the crumbled reality space results
definitively in $M_{4}$ tangent objects with spinor structure for 1-class
and 3-class particles and in $E^{4}$ tangent objects for 2-class particles
(appendix C). \ Supersymmetric particles are excluded in our hidden charge
model (section 4.1).\newline
\newline

II. The vacuum elements for the filter construction are not eigensolutions
of an independent field. This excludes Higgs particles that are, by the way,
also not necessary for mass generation by filters of our model (section 3.2).%
\newline
\newline

III. The lepton capture reaction by the 3-class particle (confinon) does not
allow a quark gluon plasma at high temperature, we find three-piece animals
also there and no independent quarks (section 5.4).\newline
\newline

IV. Dark matter particles are identified as neutral ($m,m$) 2-class
eigensolutions (table 1). The simplest dark matter particle with (1,1)
charge tori gets a mass of order 100 GeV from the filter construction (end
of section 2.3).\newline
\newline

V. The lepton capture of our model (alternatively to gauge theory
extrapolation for conventional GUT) allows a cold cosmology \cite{second
part} that is limited by existential stability to energies of order 1000
GeV. [The quarks as captured leptons introduce damping factors like equation
(\ref{Eq.5.6}) in the cage, the Hopf fiber can be borrowed from $\varphi
_{i} $ coordinates of the confinons or, for baryons, from the prebaryon.]%
\newline
\newline

\section*{7. Appendices}

\subsection*{A. Biharmonic coordinates on $S^{3}$}

Biharmonic coordinates distinguish by ($\varphi _{1},\varphi _{2}$) the two
equivalent Heegaard tori of $S^{3}$ that are used for a mathematical
characterization of the basic charge symmetry $C$ according to equation (\ref%
{Eq.3.1}) of the text (figure 7). Its conformal ''wave metric'' $g_{ik}$
corresponds to the Haar metric of the group $U(2)=(S^{1}\times S^{3})/Z_{2}$%
. Minkowski signature is allowed, because the Euler number is zero.

\begin{figure}[here]
\begin{center}
\includegraphics[width=0.5\linewidth]{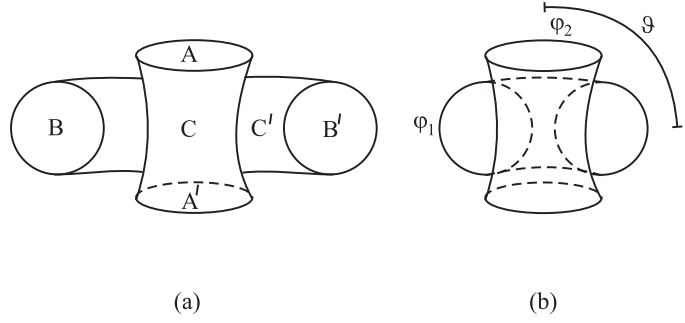}
\caption{Heegaard $D_{2}\times S^{1}$ tori of the
three-dimensional sphere $S^{3}$. (a) Stick-together model,
A~=~A$^{\prime }$, B~=~B$^{\prime } $, C~=~C$^{\prime }$. (b)
Interpenetrating Heegaard tori to get a complete covering of
$S^{3}$ by each torus, so far as possible without destroying the
topology. \label{fig7}}
\end{center}
\end{figure}

\begin{equation}
g_{ik}=\text{diag (1 , }-\cos ^{2}\vartheta \text{ , }-\sin
^{2}\vartheta \text{, }-1\text{),} \label{Eq.7.1}
\end{equation}

\begin{equation}
ds^{2}=d\tau ^{2}-(\cos ^{2}\vartheta \,d\varphi _{1}^{2}+\sin
^{2}\vartheta \,d\varphi _{2}^{2}+d\vartheta ^{2}), \label{Eq.7.2}
\end{equation}

\begin{equation}
\sqrt{-g}=\cos \vartheta \cdot \sin \vartheta . \label{Eq.7.3}
\end{equation}%
The Hopf map $S^{3}(\varphi _{1},\varphi _{2},\vartheta )\longmapsto
S^{2}(\phi ,\vartheta )$ uses two cards each of which can be connected with
one Heegaard torus,

\begin{eqnarray}
2\vartheta &=&\theta \text{ , }\varphi _{1}=0\text{ , }\varphi _{2}=-\psi
=+\phi \text{ for }S^{2}\backslash \{S\}, \label{Eq.7.4} \\
2\vartheta &=&\theta \text{ , }\varphi _{2}=0\text{ , }\varphi _{1}=+\psi
=+\phi \text{ for }S^{2}\backslash \{N\};  \nonumber
\end{eqnarray}%
the corresponding fibers are $\psi =\varphi _{1}-\varphi _{2}$ or $\varphi
=\varphi _{1}+\varphi _{2}$.

\subsection*{B. Maxwell equations for potentials on $S^{1}\times S^{3}$}

We start formally with an arbitrary metric $g^{ik}$ (Einstein's sum
convention)

\begin{equation}
(-g)^{-1/2}(\partial /\partial
x^{i})((-g)^{1/2}g^{kl}g^{im}(A_{l,m}-A_{m,l}))=0 \label{Eq.7.5a}
\end{equation}%
with the comma for differentiation of the potential components,

\begin{equation}
A_{l,m}=\partial A_{l}/\partial x^{m}. \label{Eq.7.5b}
\end{equation}%
The non-torus component is excluded. The transversality is defined by a zero
divergence. (This also excludes any charge on the hidden space),

\begin{equation}
A_{\vartheta }\equiv 0\text{ , div}_{4}A=\frac{1}{\sqrt{-g}}\frac{\partial }{%
\partial x^{i}}(\sqrt{-g}g^{ki}A_{k})=0. \label{Eq.7.6}
\end{equation}%
Explicitely,

\[
0=-\frac{1}{\cos ^{2}\vartheta }\cdot \frac{\partial }{\partial \varphi _{1}}%
\left( \frac{\partial }{\partial \varphi _{1}}A_{0}-\frac{\partial }{%
\partial \tau }A_{1}\right) -\frac{1}{\sin ^{2}\vartheta }\cdot \frac{%
\partial }{\partial \varphi _{2}}\left( \frac{\partial }{\partial \varphi
_{2}}A_{0}-\frac{\partial }{\partial \tau }A_{2}\right) -
\]

\begin{equation}
\qquad \qquad \qquad \qquad \qquad -\frac{1}{\sin \vartheta \cos \vartheta }%
\frac{\partial }{\partial \vartheta }\left( \sin \vartheta \cos \vartheta
\left( \frac{\partial }{\partial \vartheta }A_{0}\right) \right) ,
 \label{Eq.7.7}
\end{equation}

\[
0=-\frac 1{\cos ^2\vartheta }\cdot \frac \partial {\partial \tau }\left(
\frac \partial {\partial \varphi _1}A_0-\frac \partial {\partial \tau }%
A_1\right) +\frac 1{\cos ^2\vartheta }\cdot \frac 1{\sin ^2\vartheta }\frac %
\partial {\partial \varphi _2}\left( \frac \partial {\partial \varphi _1}A_2-%
\frac \partial {\partial \varphi _2}A_1\right) -
\]

\begin{equation}
\qquad \qquad \quad -\frac{1}{\sin \vartheta \cos \vartheta }\cdot \frac{%
\partial }{\partial \vartheta }\left( \frac{\sin \vartheta }{\cos \vartheta }%
\left( \frac{\partial }{\partial \vartheta }A_{1}\right) \right) ,
 \label{Eq.7.8}
\end{equation}

\[
0=-\frac 1{\sin ^2\vartheta }\cdot \frac \partial {\partial \tau }\left(
\frac \partial {\partial \varphi _2}A_0-\frac \partial {\partial \tau }%
A_2\right) +\frac 1{\sin ^2\vartheta }\cdot \frac 1{\cos ^2\vartheta }\cdot
\frac \partial {\partial \varphi _1}\left( \frac \partial {\partial \varphi
_2}A_1-\frac \partial {\partial \varphi _1}A_2\right) -
\]

\begin{equation}
\qquad \qquad -\frac{1}{\sin \vartheta \cos \vartheta }\cdot
\frac{\partial }{\partial \vartheta }\left( \frac{\cos \vartheta
}{\sin \vartheta }\left( \frac{\partial }{\partial \vartheta
}A_{2}\right) \right) , \label{Eq.7.9}
\end{equation}

\begin{equation}
0=-\frac{\partial }{\partial \tau }\left( \frac{\partial }{\partial
\vartheta }A_{0}\right) +\frac{1}{\cos \vartheta }\cdot \frac{\partial }{%
\partial \varphi _{1}}\left( \frac{\partial }{\partial \vartheta }%
A_{1}\right) +\frac{1}{\sin ^{2}\vartheta }\frac{\partial
}{\partial \varphi _{2}}\left( \frac{\partial }{\partial \vartheta
}A_{2}\right) , \label{Eq.7.10}
\end{equation}%
\begin{equation}
0=\text{ div}_{4}A=\frac{\partial }{\partial \tau
}A_{0}-\frac{1}{\cos ^{2}\vartheta }\frac{\partial }{\partial
\varphi _{1}}A_{1}-\frac{1}{\sin ^{2}\vartheta }\frac{\partial
}{\partial \varphi _{2}}A_{2}. \label{Eq.7.11}
\end{equation}%
Not only the wave metric (\ref{Eq.7.1})-(\ref{Eq.7.3}) in angles, but also
the equations (\ref{Eq.7.7})-(\ref{Eq.7.11}) are conformally invariant \cite%
{Busse1998}. The 3-class broken 4/3 charges of the confinon follow from a
decoupling scheme \cite{Busse1998} for the eigensolution of the $A_{0}$
amplitude,

\begin{equation}
A_{0}(\vartheta )=(\sin \vartheta \cos \vartheta )^{-4/3}\text{ ,
}\omega ^{2}=m_{1}^{2}=m_{2}^{2}=16/9. \label{Eq.7.12}
\end{equation}

\subsection*{C. Tangent objects and their methodical physical stratification}

As mentioned in section 2.2, the construction of an underlying tangent
object around any particle is necessary for the geometric location of points
$x$ in $\psi (x)$, which $x$ values are connected by the Feynman fabric of
figure 3. This construction \cite{Donth1988,DonthWZ89} has a physical input
from the use of physical coordinates. Different physical aspects may then
methodically be used for some formal stratification. The effects in the
reality $M$ space can then be sorted into such layers. Formal means that the
tangent object remains an indivisible whole.\newline
\newline

The {\it biquaternion layer} is the algebraic basis for the tangent objects.
They are constructed from a van der Waerden (''vdW'') algebraic
equivalence-class construction. Examples: $uu$ with no additional condition
gives tensors, $uu=0$, $uv+vu=0$ gives Grassmannians, and so on up to the
Clifford algebra used here.\newline
\newline

We start from the coordinate vectors $u_{i}$ in the hidden $H$ space (i.e.
from its ''tangent objects'' = tangent space in $H$),

\begin{equation}
u_{1}=\partial /\partial \varphi _{1}\text{ , }u_{2}=\partial /\partial
\varphi _{2}\text{ , }u_{3}=\partial /\partial \vartheta \text{ , }u_{0}%
\text{ or }u_{4}=\partial /\partial \tau . \label{Eq.7.13}
\end{equation}%
To preserve the complete charge symmetry C (\ref{Eq.3.1}) and the wave
metric signature from equations (\ref{Eq.7.1})-(\ref{Eq.7.3}), we use a \
bilinear two-$u$ construction: one $u$ from the $\varphi _{1}$ Heegaard
torus and the other $u(=v)$ from $\varphi _{2}$ torus (figure 7):

\begin{equation}
uv+vu=B(u,v)=Q(uv)-Q(u)-Q(v) \label{Eq.7.14}
\end{equation}%
with restriction to the quadratic form $Q(u)$,

\begin{equation}
u_{i}u_{i}=q_{i}\text{ , }u_{i}u_{j}+u_{j}u_{i}=q_{ij}=\text{diag (}-,-,-,+).
 \label{Eq.7.15}
\end{equation}

\begin{table}[tbp] \centering%
\caption{Formal stratification of the tangent objects\label{Tab.2}}%
\begin{tabular}{lll}
Layer name & Formal & physical \\
& specification & specification FN$^{\text{a}}$ \\ \hline
5. Mach's principle & Dollhouse universum & Gravitation. \\
\ \ \ layer & from spots of galaxies & flatness vs. \\
& by a map from & dark energy \\
& large filter elements &  \\ \hline
4. Partial chirality & Double hidden & Baryon \\
\ \ \ layer & asymmetry & survival \\ \hline
3. Exchange boson & Bilinear boson & Torus handling, \\
\ \ \ layer & construction, & Brauer Weyl \\
&  & theorem, Feynman \\
&  & diagrams \\ \hline
2. $M_{4}/E^{4}$ layer; & Tangent objects & Physical \\
\ \ Spinor $M_{4}$ space & for different & coordinates \\
\ \ or $E^{4}$ space. & particles & from nonzero \\
&  & potential components, \\
&  & wave functions \\
&  & from $H$ potentials \\ \hline
1. Biquaternion & Tangent object & Charge symmetry \\
\ \ \ layer & from biquaternions & of the hidden \\
&  & space%
\end{tabular}%
FN$^{\text{a}}$ Examples and details in the text%
\end{table}%
\newpage and $B=0$. Excluding all odd combinations of the algebra (also
because of charge symmetry $C$) gives the Clifford algebra $C_{+}$. This
algebra partitions into two biquaternions, written by complex-number
indicators ($l,i,k$) as

\begin{equation}
C_{+}=\{{\Bbb H},\widetilde{{\Bbb H}}\}\text{ , }{\Bbb
H}=\{1,l,i,k\}\text{ , }\widetilde{{\Bbb H}}=u_{5}{\Bbb H}={\Bbb
H}u_{5} \label{Eq.7.16}
\end{equation}%
with $u_{5}=u_{1}u_{2}u_{3}u_{4}$. Using physical coordinates, we get for
1-class eigensolutions exactly the vdW spinor spaces defining a flat
Minkowski structure $M_{4}$ on the tangent object (see the next layer).%
\newline
\newline

The physics of the $M_{4}$ tangent object is of electromagnetic nature, but
cannot be reduced to the photon properties alone. Instead, the above
equivalence-class construction uses three physical concepts: (a) the
signature (\ref{Eq.7.15}) left from the orthogonality of the conformal angle
(wave) metric in the $H$ space (\ref{Eq.7.1}) - (\ref{Eq.7.3}); (b) the
antisymmetry (\ref{Eq.7.15}) as for all forms in the $H$ space; and (c) the
bilinearity from the two-$u$ construction emerging from the charge symmetry $%
C$ of equation (\ref{Eq.3.1}). [Using the next layer aspects, the main point
can be worded as: the photon product-construction, suggested in the fabric
by equation (\ref{Eq.7.20}), is quenched in the tangent object by the
orthogonality $u_{1}u_{2}\sim 1$ of equation (\ref{Eq.7.18}).]\newline
\newline

In the $M_{4}/E^{4}${\it \ layer}, we introduce {\it physical coordinates} \{%
$\beta _{i}$\}. They are defined as the coordinates that carry potential
components (e.g. $\varphi _{1}$ for $A_{1}\neq 0$). The general purpose is
to give the wave function $\psi $ in the $M$ space a physical basis by the $%
A_{i}$ potentials in the $H$ space \cite{Donth1988,DonthWZ89}. Physics is
mainly ascribed to these \{$\beta _{i}$\}. The tangent object is then
specific to the class of eigensolutions (table 1, above). It is required
that the physical coordinates are linearly transferred from the tangent
vectors of the hidden $H$ space ($u_{j}$ of equation (\ref{Eq.7.13})) to the
quaternion coordinates of the tangent objects of the crumbled $M$ space.
This $H\longmapsto M$ transfer is called linear $\beta $ transfer. {\it %
Example 1} (1-class solution). The physical coordinates are called $\beta
_{i}$, the other are designed by 1, which means no transfer of physical
aspects. We have in the leptonic sector (allowing e.g. pair creation)

\begin{equation}
\beta _{1}\sim A_{1}\text{ , }\beta _{2}\sim A_{2}\text{ , }\beta _{3}\sim 1%
\text{ , }\beta _{4}\sim 1. \label{Eq.7.17}
\end{equation}%
Orthogonality of the biharmonic coordinates is then expressed by

\begin{equation}
u_{1}u_{2}\sim \left\{
\begin{array}{ccc}
\beta _{1} & \text{for} & \oplus \text{ 1-class solutions }\psi ^{(1)} \\
\beta _{2} & \text{for} & \ominus \text{ 1-class solutions }\psi ^{(2)} \\
1 & \text{for} & \text{products like a photon }A.%
\end{array}%
\right. \label{Eq.7.18}
\end{equation}%
This yields, as mentioned above, the two nontrivial vdW spinor spaces giving
the usual bilinear Minkowski $M_{4}$ structure for the $M_{4}$ tangent
objects for 1-class particles.\newline
\newline

{\it Example 2 }(2-class solution). The linear $\beta $ transfer results
here in a situation, that the physical coordinates can be collected in one
quaternion, e.g. ${\Bbb H}$; we do not obtain two nontrivial spinor spaces.
In \cite{DonthWZ89} is shown: If only one quaternion component (i.e. the
quaternion ${\Bbb H}$ or $\widetilde{{\Bbb H}}$) is physically relevant for
the $M$ space tangent object, then its total biquaternion metric is
Euclidean, $E^{4}$. This leads to an elliptic ''wave equation'' for
non-isolated dark matter particles and prebaryons. (For isolated objects
with inside $E^{4}/M_{4}$ crumbling, the outside is Minkowskian flat, only $%
M_{4}$.)\newline
\newline

In the {\it exchange boson layer}, the bosons are constructed by the Feynman
fabric of figure 3. The electroweak exchange bosons are formed from the
bilinear charge-conserving construction

\begin{equation}
(A,\text{ }W^{\pm },\text{ }Z)\sim (A_{1}\text{ type})\ast (A_{2}\text{ type}%
). \label{Eq.7.19}
\end{equation}%
The fabric is reflected in the tangent object by the $\ast $ which
symbolizes a junction between the two spinor spaces. We obtain

\begin{equation}
A=%
{0 \choose v}%
\ast
{\bar{0} \choose v}%
\text{ , }Z=%
{+ \choose v}%
\ast
{- \choose v}%
\text{ , }W^{+}=%
{+ \choose v}%
\ast
{\bar{0} \choose v}%
\text{ , }W^{-}=%
{0 \choose v}%
\ast
{- \choose v}%
, \label{Eq.7.20}
\end{equation}%
where the (0,$\bar{0}$) defines the photon particle by the $\tau $ torus, ($%
+,-$) are the two electrical charges, and $v$ is some indication of the
vacuum filters for the bosons. Equations (\ref{Eq.7.20}) are the linear $%
\beta $-transfer roots for the conventional Brauer-Weyl theorem

\begin{equation}
\gamma ^{\mu }\partial _{\mu }\rightarrow \{\gamma ^{\mu }\partial
_{\mu },\gamma ^{\mu }A_{\mu },\gamma ^{\mu }Z_{\mu },\gamma ^{\mu
}W_{\mu }^{\pm }\}. \label{Eq.7.21}
\end{equation}%
Remark: The relation of these boson construction to the vectors (operators)
in the hidden charge is sketched in table 3. The ${\bf L}$ symbol means a so
called torus exchange vector, constructed from the ($L_{1},L_{2}$) invariant
vector fields of appendix D,

\begin{equation}
{L_{1} \choose L_{2}}%
=%
{\cos \varphi \quad \sin \varphi  \choose -\sin \varphi \quad \cos \varphi }%
{{\bf L} \choose \partial _{3}}%
, \label{Eq.7.22}
\end{equation}

\begin{table}[tbp] \centering%
\caption{Torus content and torus handling for electroweak exchange
bosons\label{Tab.3}}%
\begin{tabular}{lll}
& Charge torus & Torus induction \\
Particle & content & in $M_{4}$ Lagrangian \\
& ($m=1$) & (section 4) FN$^{\text{a}}$ \\ \hline
$A$ & no & ${\bf L}A=0$ \\
$Z$ & two: $e^{im\varphi _{1}},e^{-im\varphi _{2}}$ & ${\bf L}Z=\cot
\vartheta \cdot \partial _{1}-\tan \vartheta \cdot \partial _{2}$ \\
$W^{+}$ & $e^{-im\varphi _{2}}$ & ${\bf L}W^{+}=(\partial _{1}=0,-\tan
\vartheta \cdot \partial _{2})$ \\
$W^{-}$ & $e^{+im\varphi _{1}}$ & ${\bf L}W^{-}=(\cot \vartheta \cdot
\partial _{1},\partial _{2}=0)$ \\ \hline
$\nu _{e},\nu _{\mu },\nu _{\tau }$ & no & ${\bf L}\nu =0$ \\
$e,\mu ,\tau $ & $e^{im\varphi _{1}}$ & ${\bf L}e^{-}=(\cot \vartheta \cdot
\partial _{1},\partial _{2}=0)$ etc. \\
$\bar{e},\bar{\mu},\bar{\tau}$ & $e^{-im\varphi _{2}}$ & ${\bf L}%
e^{+}=(\partial _{1}=0,-\tan \vartheta \cdot \partial _{2})$ etc.%
\end{tabular}%
FN$^{\text{a}}$: see equations (\ref{Eq.7.22}-\ref{Eq.7.23}) and (\ref%
{Eq.4.2}).%
\end{table}%

\begin{equation}
{\bf L}=\cos \vartheta \cdot \partial _{1}-\tan \vartheta \cdot \partial _{2}%
\text{ , }\partial =%
{\sin \varphi  \choose \cos \varphi }%
\partial _{3}, \label{Eq.7.23}
\end{equation}%
with $\varphi $ the fiber coordinate of the Hopf map for the $L_{i}$ vector
fields and $\partial $ the mass operator (\ref{Eq.4.2}). Torus handling
means the knowledge how to operate with the tori of the hidden charge to get
the electroweak Lagrangian and the Feynman diagrams in the reality space.
Torus induction is a part of this handling: which differentiations of (\ref%
{Eq.7.23}) generate the exchange bosons.\newline

In the {\it partial chirality layer}, a twofold hidden charge asymmetry is
formulated. Twofold means hidden in the $H$ space and hidden behind the
charge symmetry $C$ by means of the $C_{+}$ algebra. In this hypothetical
layer \cite{Donth2006} we ask how the physical coordinates and the wave
functions react on the charge exchange $C$: $u_{1}\leftrightarrow u_{2}$.
This chirality is defined by hidden nontrivialities with respect to the
Heegaard tori, expressed by all possible $u_{i}$ combination (including also
$u_{1}$, $u_{2}$, and $u_{1}u_{2}u_{3}$), e.g. by the differences between
the \{$u_{1},u_{1}u_{2},u_{1}u_{2}u_{3}$\} components. This chirality is
e.g. different for 1-class spinors in $M_{4}$ and constructions for 2-class
particles in $E^{4}$ (prebaryons, baryon charge, dark matter). Partial
chirality cannot be defined on the basis of a $M_{4}$ manifold. The partial
chirality ($P^{\prime }$) is conserved for 1-class and broken for 2-class
particles. Requiring $P^{\prime }CT=1$, then $P^{\prime }$ properties can be
transferred to time reversal violation $T$. The participation of the
prebaryon in the lepton capture reaction (\ref{Eq.5.10}) for baryons can
qualitatively explain their cosmological surviving \cite{Donth2006}.\newline
\newline

In the {\it Mach's principle layer}, a map of galaxies from the whole
universe into spots on each tangent object (dollhouse universe) is
considered, that is mediated by the large filter elements for the
culminating-point particle in the center of any tangent object (details see %
\cite{second part}). The corresponding elements of the reality space are
specific with respect to number and ''direction'' of gravitons that take
share. These species define an additional shadow metric $g(x)$ over an
otherwise flat Minkowski space from isolated $M_{4}$ tangent objects without
a metric diffeomorphism.

\subsection*{D. $u(2)$ invariant vector fields}

These fields transport the local $u(2)$ symmetry over the whole $U(2)$ group
space, and therefore over the whole hidden $S^{1}\times S^{3}$ space because
of the $U(2)=(S^{1}\times S^{3})/Z_{2}$ covering. These fields correspond to
the ''rotations'' of figure 2 that are assumed to induce motions in the
reality space defining the electroweak Lagrangian in the context of Feynman
connections. There are 7 different fields, four left ones ($L_{1},...,L_{4}$%
) and four right ones ($R_{1},...,R_{4}$) with $L_{4}=R_{4}$. (Such fields
are called Killing vectors in general relativity.) We restrict ourselves to
the left ones (left invariant multiplication) and have the following four
fields, corresponding to the $u(2)$ matrices as indicated:

\begin{equation}
\left(
\begin{array}{cc}
0 & i \\
i & 0%
\end{array}%
\right) \text{: }L_{1}=\cos \varphi \cdot \cot \vartheta \cdot \partial
/\partial \varphi _{1}-\cos \varphi \cdot \tan \vartheta \cdot \partial
/\partial \varphi _{2}+\sin \varphi \cdot \partial /\partial \vartheta ,
 \label{Eq.7.24}
\end{equation}

\begin{equation}
\left(
\begin{array}{cc}
0 & 1 \\
-1 & 0%
\end{array}%
\right) \text{: }L_{2}=-\sin \varphi \cdot \cot \vartheta \cdot \partial
/\partial \varphi _{1}+\sin \varphi \cdot \tan \vartheta \cdot \partial
/\partial \varphi _{2}+\cos \varphi \cdot \partial /\partial \vartheta ,
 \label{Eq.7.25}
\end{equation}

\begin{equation}
\left(
\begin{array}{cc}
i & 0 \\
0 & -i%
\end{array}%
\right) \text{: }L_{3}=-\partial /\partial \varphi _{1}-\partial
/\partial \varphi _{2}, \label{Eq.7.26}
\end{equation}

\begin{equation}
\left(
\begin{array}{cc}
0 & i \\
i & 0%
\end{array}%
\right) :L_{4}=\partial /\partial \tau , \label{Eq.7.2x}
\end{equation}%
with the Hopf map fiber $\varphi =\varphi _{1}+\varphi _{2}$.

\subsection*{E. Empirical electroweak Lagrangian}

The conventional Lagrangian is written in a form that especially well fits
to a comparison with the above vector fields:

\begin{eqnarray}
L_{C}/e_{0} &=&\bar{e}\gamma _{\mu }eA_{\mu }\qquad \qquad \qquad
\qquad \qquad \qquad \qquad \qquad \qquad \text{em current}
\label{Eq.7.28} \\
&&+\frac{1}{2}\{(\tan \theta _{W}-\cot \theta _{W})\bar{e}_{L}\gamma _{\mu
}e_{L}\qquad \qquad \qquad \quad \ \ \ \ \text{neutral }e_{L}\text{ current}
 \label{Eq.7.29} \\
&&+(\tan \theta _{W}+\cot \theta _{W})\bar{\nu}\gamma _{\mu }\nu \qquad
\qquad \qquad \qquad \qquad \ \text{neutral }\nu \text{ current}
 \label{Eq.7.30} \\
&&+(\tan \theta _{W}+\tan \theta _{W})\bar{e}_{R}\gamma _{\mu }e_{R}\}Z_{\mu
}\qquad \qquad \qquad \quad \ \ \text{neutral }e_{R}\text{ current }
 \label{Eq.7.31} \\
&&+\text{cosec}\theta _{W}(2\sqrt{2})^{-1}\bar{\nu}\gamma _{\mu }(1+\gamma
_{5})eW_{\mu }^{+}+h.c.\qquad \ \ \ \ \ \ \text{charged current}
 \label{Eq.7.32} \\
&&+\frac{1}{2}\bar{\nu}\gamma _{\mu }\partial _{\mu }(1+\gamma _{5})\nu +%
\bar{e}\gamma _{\mu }\partial _{\mu }e\qquad \qquad \qquad \ \ \ \ \ \ \ \ \
\text{kinetics} \label{Eq.7.33} \\
&&+h_{e}%
{\bar{\nu} \choose e_{L}}%
\phi e_{R}+...\text{ \ \ \ \ \ \ \ \ \ \ \ \ \ \ \ \ \ \ \ \ \ \ \
\ \ \ \ \ \ \ \ \ \ \ \ \ \ \ \ mass term for }e. \label{Eq.7.34}
\end{eqnarray}%
with $h_{e}$... the adjustable Yukawa mass parameters and $\phi $
the Higgs field.

\section*{Acknowledgements}

This work was partly supported by the DFG Sonderforschungsbereich SFB 418
and by the Fonds Chemische Industrie, FCI. The author thanks in particular
Steffen Trimper, the speaker of the SFB, and has profited from stimulating
discussions with Karsten Busse and Michael Schulz, and from some
calculations by Jens-Uwe Sommer.

\end{document}